\documentstyle[12lomcon,graphicx, cite]{article}
\begin{document}

\title{Searching for neutrino oscillations with OPERA}

\author{ N. Savvinov \footnote{e-mail: nikolay.savvinov@lhep.unibe.ch} (on behalf of the OPERA collaboration)}

\address{Laboratory for High Energy Physics, University of Bern, CH-3012, Switzerland}


\maketitle\abstracts{ The OPERA experiment will search for
neutrino oscillations using a muon neutrino beam and a hybrid
emulsion-scintillator detector. Basic principles, current status
and expected performance of the experiment are discussed.}

\section{Motivation}

Results from Super-Kamiokande \cite{sk} and K2K \cite{k2k}
together with those of CHOOZ \cite{chooz} gained strong evidence
in favor of the $\nu_{\mu} \rightarrow \nu_\tau$ scenario for the
atmospheric neutrino anomaly. OPERA intends to confirm this result
by observing the appearance of $\nu_\tau$ in a $\nu_\mu$ beam.
As a byproduct of the
experiment, a significant improvement of the current CHOOZ
limit~\cite{chooz} on $\theta_{13}$ is expected.

\section{Experimental strategy}


\begin{figure}[p]
\begin{center}
\includegraphics[width=0.9\textwidth]{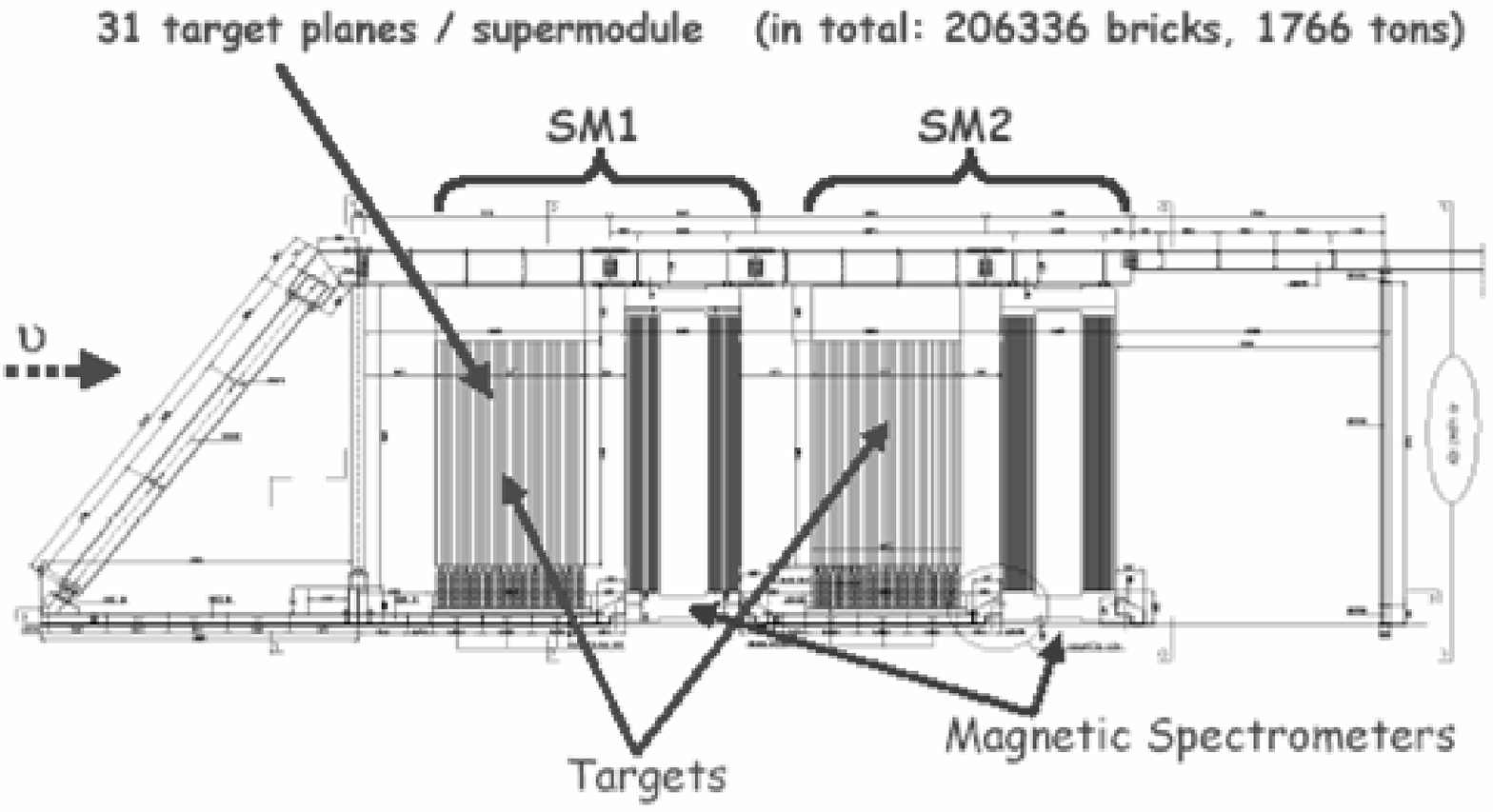}
\end{center}
\caption{Schematic view of the OPERA detector.}
\label{fig:opera_detector}
\begin{center}
\end{center}
\end{figure}

OPERA will use the CNGS (CERN Neutrinos to Gran Sasso) $\nu_\mu$
beam.
The detector will be placed in the Gran Sasso underground
laboratory in Italy, 735~km away from the source.

The detection of $\nu_\tau$ will be based on direct observation of
$\tau$ decay topologies.
 Since $\tau$ decay length under experimental conditions is very
short (of order of 1~mm), a detector with a very high spatial
resolution is required. This will be accomplished by using nuclear
emulsions.


The OPERA emulsions films consist of two 40~$\mu$m emulsion layers
separated by a 200~$\mu$m  plastic base. The emulsions are
sandwiched with 1~mm lead sheets, according to the so-called
emulsion cloud chamber (ECC) technique. The basic unit of the
OPERA detector, a brick, contains 57 emulsion films and 56 lead
plates. The brick dimensions are 10.2 $\times$ 12.7 $\times$ 7.5
cm and the weight is 8.6~kg. The overall weight of the OPERA
detector will be 1.7~kton.

In order to identify the location of the bricks containing the
neutrino interaction point, brick layers are interspaced with
electronic detectors. The Target Tracker (TT) planes consist of
horizontal and vertical plastic scintillator strips read out by
64-channel PMTs. The DAQ of TT defines the trigger for the brick
extraction for analysis.

The OPERA detector will be organized into two large sections
("supermodules") of 31 brick and TT planes. Each supermodule will
be followed by a muon spectrometer for $\mu$ identification,
momentum and charge measurement and charm background reduction.
Each muon spectrometer is composed of a 1.55~T dipolar magnet, 22
RPC layers and 6 sections of drift tubes.

Once the interaction candidate brick(s) is (are) identified, its
(their) extraction will be handled by two designated robots called
BMS ("brick manipulation system"). After extraction, the brick is
exposed to cosmic rays for film alignment, and then disassembled.
The emulsions are developed on-site and then sent to participating
laboratories for scanning. After scanning, particle tracks are
analyzed and the event is reconstructed using momentum information
from multiple Coulomb scattering and readouts of the TT and the
muon spectrometers.







\section{Installation of the experiment}

\subsection{Neutrino Beam}

\begin{figure}[p]
\begin{center}
\includegraphics[width=0.9\textwidth]{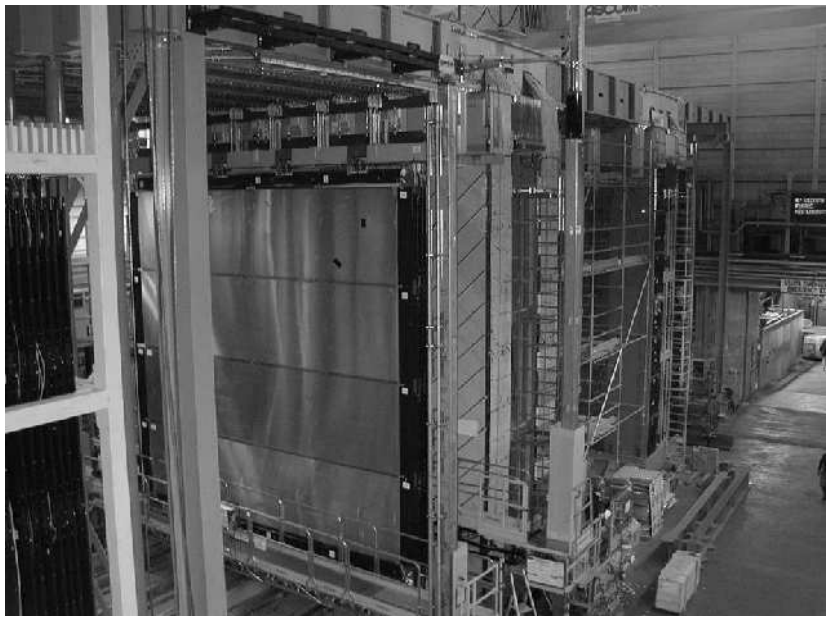}
\end{center}
\caption{Construction of the OPERA detector (June 2005). The
neutrino beam will come from the left.} \label{fig:opera_detector}
\end{figure}

The CNGS project has successfully completed the civil engineering
stage as well as the installation of hadron shop, decay tube and
general services. Currently, installation of the proton beamline,
target, horn, deflector and shielding is underway. The
comissioning is expected in Spring 2006.

\subsection{Spectrometers}

The magnets for both muon spectrometers are now in place.
Installation of detectors for the spectrometers is in progress.
The 4 downstream RPC planes of the first supermodule are already
taking cosmics data.

\subsection{Target modules}


All required TT modules have been produced and are now delivered
to Gran Sasso on a ready-to-install basis. As of September 2005,
about 2/3 of TT and brick walls for the first supermodule are
installed. It is expected that by November 2005 all TT
installations for the first supermodule will be completed. The
brick assembly machine (BAM) is expected to be fully commissioned
by December 2005 and should start producing lead-emulsion bricks
in January 2006 . Brick filling of the first supermodule should
start in January 2006 and continue until the expected CNGS beam
arrival in July 2006. After that, while taking data with one
supermodule, the collaboration will continue to fill the second
supermodule with bricks.

\section{Emulsion scanning}

\begin{figure}[th]
\caption{Few sample reconstructed events from a 10~GeV pion test
beam.The beam is entering the picture from the right.}
\label{fig:pions}
\begin{center}
\includegraphics[width=0.75\textwidth]{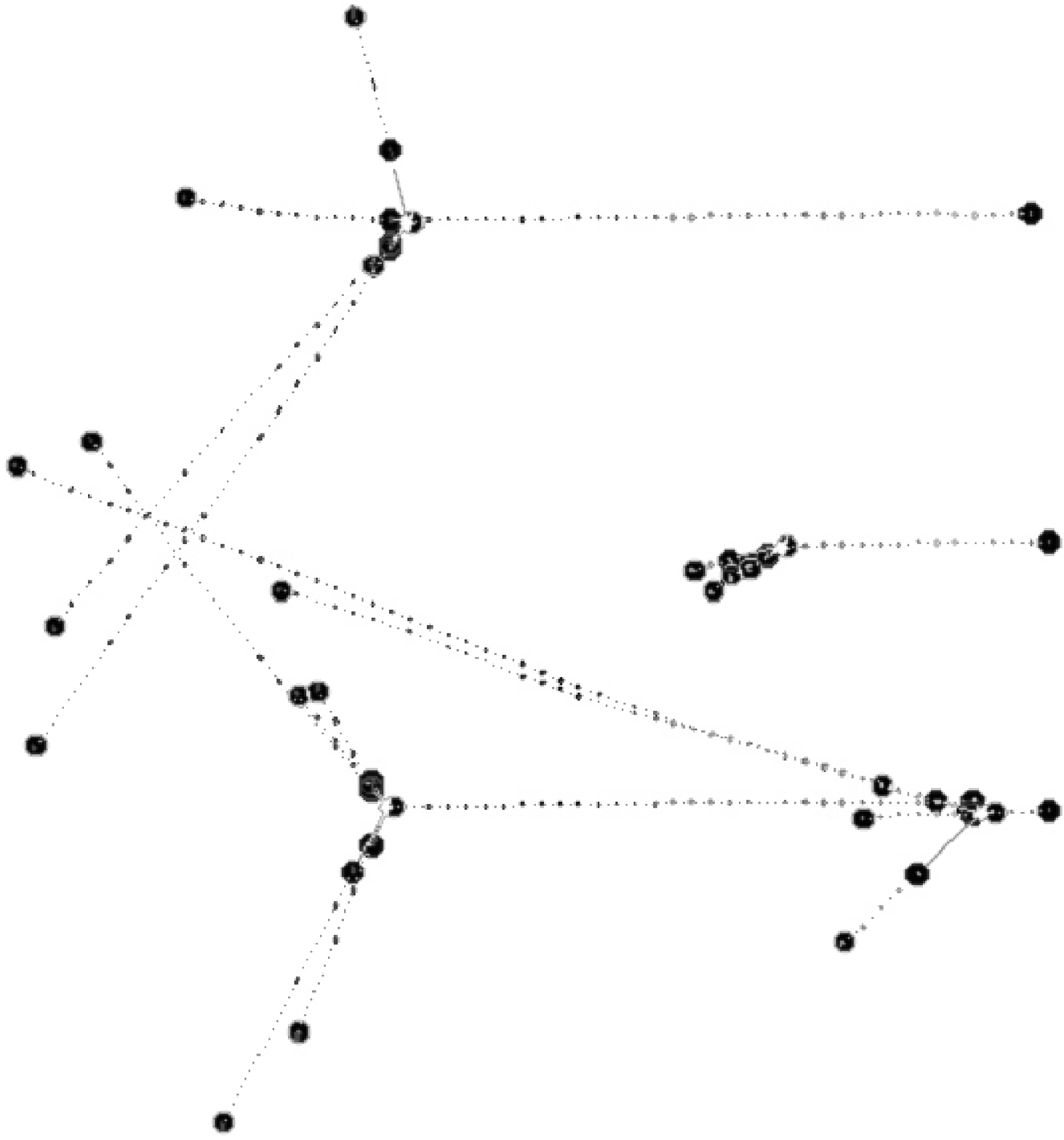}
\end{center}
\end{figure}

The OPERA scanning system uses many ideas and approaches that
originated during development of the automatic scanning system for
\nobreak{CHORUS}. The total surface of OPERA emulsions is,
however, more than 170,000~m$^2$, as compared to 500 m$^2$ for
CHORUS~\cite{chorus}. Such a high scanning load requires
innovations in both scanning hardware and analysis algorithms,
which have been successfully implemented in OPERA
computer-controlled scanning stations.

Thanks to these innovations, the scanning speed has been increased
more than by a factor of 20 with respect to the previous systems
and reached over 20~cm$^2$ per hour. Online 3D reconstruction of
particle tracks is performed with a sub-micron precision and
efficiency better than~95\%.

Scanning systems installed in Europe and Japan are now operational
and are analyzing data from various test beams. A sample
reconstructed event from a CERN 10~GeV pion test beam is shown in
Figure~\ref{fig:pions}.

The overall scanning capacity of the collaboration is estimated
around 30~bricks per day, which fulfills the requirements of the
experiment.


\section{Expected event and background rates}


The overall efficiency of $\tau$-decay observation in OPERA was
calculated to be 9.1\%. This figure is based on both short and
long decays of $\tau \rightarrow e$, $\tau \rightarrow \mu$ and
$\tau \rightarrow h$ channels. A possibility of inclusion of the
$\tau \rightarrow 3h$ channel with potential improvement in
efficiency by about 1\% is being investigated. Finally, a recent
study shows a possibility of a 10~\% improvement in brick finding
efficiency.

The event rate of OPERA strongly depends on the oscillation
parameters. Calculations for several values of $\Delta m^2_{atm}$
(corresponding to the central value of Super-Kamiokande best fit
and the limits of the 90\% confidence interval) are presented in
Table~\ref{tbl:number_signal_bg}.

The main sources of background are charm decays, hadron
re-interactions and large angle $\mu$ scattering. Algorithms
employing dE/dx information for improved $\pi/\mu$ separation are
currently being developed. It is expected that they would reduce
charm background by about 40\% (down to about 0.28 event). The
result of recalculation of the large angle $\mu$ scattering
background including nucleon form factors is 5 times lower than
the upper limit from the CHORUS measurement used earlier.

\begin{table}[t]
\caption{Expected number of observed signal and background events
in 5 years of running for nominal calculations and envisaged
improvements of efficiency and background reduction.}
\label{tbl:number_signal_bg}
\begin{center}
\begin{tabular}{|c|ccc|c|}
\hline
 & \multicolumn{3}{c|}{Signal} & BG \\
$\Delta m^2~(10^{-3}$eV$^2$) & 1.9 & 2.4 & 3.0 & \\
\hline
Nominal & 6.6 & 10.5 & 16.4 & 0.7 \\
Improved eff. & 8.0 & 12.8 & 19.9 & 1.0 \\
+ BG reduction  & 8.0 & 12.8 & 19.9 & 0.8  \\
\hline
\end{tabular}
\end{center}
\end{table}

\section{Conclusions \label{section:conclusions}}

The CNGS and the OPERA detector construction are progressing and
the experiment should be ready to take data in August 2006. Based
on Super-Kamiokande best fit for $\Delta m^2_{atm}$, OPERA should
see in 5 years about 12.8 $\tau$ events. The background is
expected on the level of 1 event. An ongoing effort to improve the
efficiency and to further suppress the background may improve
these figures.



\end{document}